\documentclass[12pt,preprint]{aastex}
\usepackage{multirow}
 \newcommand{\beq}{\begin{equation}}
 \newcommand{\eeq}{\end{equation}}
 \newcommand{\beqn}{\begin{eqnarray}}
 \newcommand{\eeqn}{\end{eqnarray}}

\begin{document}

 \title{\bf{
The Possible Orbital Decay and Transit Timing Variations
of the Planet WASP-43b}}
\author{Ing-Guey Jiang$^1$, Chien-Yo Lai$^1$, 
Alexander Savushkin$^2$, 
David Mkrtichian$^{3}$,\\
Kirill Antonyuk$^{2,4}$,
Evgeny Griv$^5$,
He-Feng Hsieh$^1$,
Li-Chin Yeh$^6$
 }
\affil{
 {$^1$Department of Physics and Institute of Astronomy,}\\
 {National Tsing-Hua University, Hsin-Chu, Taiwan}\\
{$^2$Crimean Astrophysical Observatory, 298409, Nauchny, Crimea}\\
{$^3$National Astronomical Research Institute of Thailand (NARIT),}\\
{Siripanich Building, 191 Huaykaew Road, Muang District, Chiangmai, Thailand}\\
{$^4$Special Astrophysical Observatory, Russian Academy of Sciences,
Nizhnii Arkhyz, Russia}\\
{$^5$Department of Physics, Ben-Gurion University, Beer-Sheva 84105, Israel}\\
{$^6$ Department of Applied Mathematics,}\\
{National Hsinchu University of Education, Hsin-Chu, Taiwan}\\
}

\begin{abstract}
Motivated by the previously reported high orbital decay rate of the 
planet WASP-43b, 
eight newly transit light curves are obtained and presented. 
Together with other data in literature, we perform 
a self-consistent timing analysis with data 
covering a timescale of 1849 epochs.
The results give an orbital decay rate
$dP/dt$ = $-0.02890795\pm 0.00772547$ sec/year, which is
one order smaller than previous values.
This slow decay rate corresponds to a normally assumed theoretical value
of stellar tidal dissipation factor.
In addition, through the frequency analysis, the 
transit timing variations presented here are unlikely to be periodic, but 
could be signals of a slow orbital decay.
\end{abstract}


\section{Introduction}

The discoveries of new extra-solar planets continue to be exciting 
and have revealed many new implications about the formation and evolution of 
planetary systems. 
Though the majority of them was detected by the method of Doppler shift
(Marcy \& Butler 1998),
other methods such as transit, direct imaging etc. also made 
impressed contributions.
Because the orbital configurations of extra-solar planetary systems 
are generally quite different from our Solar system, many investigations
on their dynamical properties and evolution have been done 
(see Jiang \& Ip 2001, Ji et al. 2002,
Jiang et al. 2003, Jiang \& Yeh 2004a,
Jiang \& Yeh 2004b, Jiang \& Yeh 2007, 
Mordasini et al. 2009). 
In addition, the further observational effort also produces
new fruitful results continuously. For example, 
Kepler space telescope has discovered many multiple planetary systems 
through the transit method (Lissauer et al. 2011).
Maciejewski et al. (2010) and Jiang et al. (2013)
discovered possible transit timing variations (TTVs) which could 
imply the existence of additional bodies in these planetary systems. 
Lee et al. (2014) also found planetary companions around evolved stars
through the method of radial velocities by Doppler shift.

Among these, 
those extreme systems with very short orbital period
have particularly raised many interesting questions 
such as where they could have 
formed, how they would have migrated to current positions, and how 
stable their current orbits are etc.
Their physical properties have also been seriously investigated with 
great effort. For example, WASP-12 planetary system,
discovered by Hebb et al. (2009), was one of the well known 
extreme systems that attracted much attention. 
The planet was argued to be losing mass by exceeding its Roche lobe.
Due to the falling of planetary gas towards the host star through the
first Lagrange point, it is likely to form an accretion disk 
(Li et al. 2010).
This might lead to the transfer of metals and thus enhance the stellar
metallicity.
Maciejewski et al. (2011) employed a high-precision photometric 
monitoring to study this system and greatly improved the determination 
of WASP-12b planetary properties.

On the other hand, the WASP-43 planetary system, first discovered by 
Hellier et al. (2011), is another case with 
an even smaller orbit. 
The planet is moving around a low mass K star with 
an orbital period only about 0.8 days. 
With a mass of 1.8 Jupiter Mass,
it is one of the most massive exoplanets carrying 
an extremely short orbital period. 

The existence of WASP-43 system has therefore triggered
the study of thermal radiation from exoplanets.
For example, Wang et al. (2013) confirmed the thermal emission 
from the planet WASP-43b.  
Chen et al.(2014) observed one transit and one occultation
event in many bands simultaneously. They detected
the day-side thermal emission in the $K$-band. 
Moreover, Kreidberg et al. (2014) 
determined the water abundance in the atmosphere of WASP-43b 
based on the observations through Hubble Space Telescope.  

As discussed in Jiang et al. (2003), 
a system with a close-in planet would experience an orbital decay through
star-planet tidal interactions.
Indeed, through the XMM-Newton observations, 
Czesla et al. (2013) showed an X-ray detection and 
claimed that WASP-43 is an active K-star, which could be related with 
tidal interactions. 
In order to obtain more precise measurements of the 
characteristics of this system,  Gillon et al. (2012)
performed an intense photometric monitoring by ground-based 
telescopes. The physical parameters have been measured 
with much higher precision. Employing their data, the 
atmosphere of WASP-43b was modeled. However, 
they concluded that their transit data presented 
no sign of transit timing variations.  

Later, through a timing analysis on the transits of 
WASP-43b, Blecic et al. (2014) 
proposed an orbital period decreasing rate  
about 0.095 second per year. With the data from 
Gran Telescopio Canarias (GTC),
Murgas et al. (2014) also claimed an orbital decay
with period decreasing rate about 0.15 second per year
and suggested that a further timing analysis over future years 
would be important. 

Motivated by the above interesting results,   
we employ two telescopes to monitor the WASP-43b transit events
and obtain eight new transit light curves.
Combining our own data  
with available published 
photometric transit data of WASP-43b, we investigate 
the possible timing variations or orbital decay here.
Since these data cover more than 1800 epochs of the orbital evolution,
our results shall serve as the most updated reference for this
system. Our observational data are described in Section 2, 
the analysis of light curves is in Section 3, the results of 
transit timing variations are presented in Section 4, 
and finally the concluding remarks
are provided in Section 5. 

\section{Observational Data}

\subsection{Observations and Data Reduction}

In this project, 
two telescopes were employed to observe
the transits of 
WASP-43b. One is the 1.25-meter telescope (AZT-11) 
at the Crimean Astrophysical Observatory (CrAO) in Nauchny, Crimea,
and another is the 60-inch telescope (P60) at Palomar Observatory 
in California, USA.
We successfully performed one complete transit observation 
with AZT-11 in 2012 and seven with P60 in 2014 and 2015.
A summary of the above observations is presented in Table 1.

After some standard procedures such as flat-field corrections etc., 
we first use the IRAF task, $daofind$, to pick bright 
stars and then the task, $phot$, to measure these stars' fluxes 
in each image. The light curves of these bright stars
are thus determined (Jiang et al. 2013, Sun et al. 2015). 
In order to decide which stars could 
be comparison stars, we first choose those with higher 
brightness consistency as candidates, i.e. candidate stars.
Therefore, we calculate the 
Pearson's correlation coefficient between 
any two of these light curves and use 0.9 as the criterion.
The candidate stars are a set of stars in which 
the correlation coefficient between any pairs of their light curves 
must be more than 0.9, in order to ensure the brightness 
consistency and that none of the candidate stars
are variable objects. 
Finally, the flux of WASP-43 is divided by any possible combination 
of the fluxes from candidate stars.
For example, when there are three candidate stars,
the flux of WASP-43 is divided by the summation
of all three candidate stars, the summation of any possible pairs 
from these candidate stars, and also the flux of individual 
candidate stars.  
Each of the above operations leads to one calibrated light curve, 
and the one with the smallest out-of-transit root-mean-square
deviation becomes the light curve of WASP-43. 
Note that the out-of-transit root-mean-square deviation
is determined after the normalization process, 
which would be described in Section 2.3 later. 
The comparison stars are those involved in the determination of 
the light curve of WASP-43.
The number of bright stars, candidate stars, and comparison stars
are listed in Table 2. 

\subsection{Other Observational Data from Literature}
In addition to our own light curves, it will be very helpful
to employ those publicly available transit data 
in previous work.  
With both our own and other transit light curves, 
we could therefore cover a large number of 
epochs for the investigation on possible transit timing variations. 
We review all WASP-43b papers and find that 
there are five papers in which the electronic files
of transit light curves are provided.

Gillon et al. (2012) employed the 60cm telescope, TRAPPIST
(TRAnsiting Planets and PlanetesImals Small Telescope), 
in the Astrodon $I+z$ filter to obtain 20 light curves
and the 1.2m Euler Swiss telescope in 
the Gunn-$r'$ filter to obtain three light curves.
Two of the above light curves are actually 
for the same transit event. That is, Epoch 38
is observed by both telescopes. Note that the epochs are given 
an identification number following the convention that the transit observed 
in Hellier et al. (2011) is Epoch 0. 
Chen et al. (2014) observed the transit event of Epoch 499 with the 
GROND instrument mounted on 2.2m MPG/ESO telescope
in seven bands: Sloan $g', r', i', z'$ and NIR $J, H, K$. 
We take the light curve in $J$ band,
because only $J, H$, and $K$ bands have the information 
of seeing and the wavelength of $J$ band is the closest to $R$ band, 
the one we used for our own observations.
Maciejewski et al. (2013) provided the light curves of Epoch 543
and Epoch 1032. 
Murgas et al. (2014) used GTC (Gran Telescopio Canarias) 
instrument OSIRIS to obtain long-slit spectra. We choose the 
white light-curve data to do the analysis in this paper.
In addition, there are
seven light curves available from Ricci et al.(2015). 

Therefore, we take 23 light curves from Gillon et al. (2012).
In addition, we get one light curve from Chen et al. (2014), 
two light curves from Maciejewski et al. (2013), 
one from Murgas et al. (2014), and seven from 
Ricci et al.(2015).
In total, we have 34 light curves
taken from published papers.

We do not simply use the mid-transit times written in these papers, 
but re-analyze all the photometric data with the same procedure 
and software to perform 
parameter fitting in a consistent way.
Because all these data go
through the same transit modeling and fitting procedure 
as our own data,
it can ensure that the results are reliable.

\subsection{The Normalization and Time Stamp of Light Curves}

For all the previously mentioned light curves, 
including eight from our work and 34 from the 
published papers, we further consider the airmass and 
seeing effects here. 
As the procedure in Murgas et al. (2014),  
a 3rd-degree polynomial is used to model the airmass effect, 
and a linear function is employed to model 
the seeing effect.
The original light curve, $F_{o}(t)$, 
can be expressed as: 
\begin{equation}
F_{o}(t) = F(t) \mathcal{P}(t) \mathcal{Q}(s),
\end{equation}
where $F(t)$ is the corrected light curve, 
$\mathcal{P}(t) = a_0 + a_1 t + a_2 t^2+ a_3 t^3$, 
and $\mathcal{Q}(s) = 1 + c_0 s$,
where $s$ is the seeing of each image. 
(Note that, in Maciejewski et al. 2013 and Ricci et al. 2015, 
the seeing is not known and no seeing correction can be done. 
We thus set $\mathcal{Q}(s)=1$ for light curves from these two papers.)  
We numerically search the best values of five parameter 
$a_0, a_1, a_2, a_3, c_0$ to make out-of-transit part
of $F(t)$   close to unity with smallest standard deviations,
and thus normalize all the light curves.
$F(t)$ would be simply called the observational 
light curves and used in any further analysis 
for the rest of this paper.  

On the other hand, the time stamp we use is the Barycentric Julian Date 
in the Barycentric Dynamical Time ($BJD_{TDB}$).
We compute the UT time of mid exposure from the recorded header 
and convert the time stamp to 
$BJD_{TDB}$ as in Eastman et al. (2010).

\section{The Analysis of Light Curves}

The Transit Analysis Package (TAP) presented by 
Gazak et al. (2012) is used to obtain transit models and 
corresponding parameters from all the above 42 light curves.  
TAP employs the light-curve models of Mandel \& Agol (2002), 
the wavelet-based likelihood function developed by Carter \& Winn (2009),
and Markov Chain Monte Carlo (MCMC) technique to
determine the parameters.

All 42 light curves are loaded into TAP  
and analyzed simultaneously. 
For each light curve, five MCMC chains of length 500,000 are computed. 
To start an MCMC chain in TAP, we need to set the initial values 
of the following parameters: 
orbital period $P$, orbital inclination $i$, semi-major axis $a$ 
(in the unit of stellar radius $R_{\ast}$), the planet's radius $R_{\rm p}$ 
(in the unit of stellar radius), 
the mid-transit time $T_{\rm m}$, the linear limb darkening 
coefficient $u_1$, the quadratic limb darkening 
coefficient $u_2$, orbital eccentricity $e$ and 
the longitude of periastron $\omega$. Once the initial 
values are set, one could choose any one of the above to be: 
(1) completely fixed (2) 
completely free to vary or (3) varying following a Gaussian function, 
i.e., Gaussian prior, during the MCMC chain in TAP.
Moreover, any of the above parameters which is not completely fixed  
can be linked among different light curves. 
The orbital period is treated as 
a fixed parameter $P$ = 0.81347753, which is taken from Table 5 of 
Gillon et al. (2012). The initial values of inclination $i$,
semi-major axis, and planet's radius are all from Gillon et al.(2012),
i.e.  $i$=82.33, $a/R_{\ast}$=4.918, and $R_{\rm p}/R_{\ast}$=0.15945.
They are completely free to vary and linked among all light curves.  
We leave the mid-transit times
$T_{\rm m}$ to be completely free during TAP runs and  
it is only linked among those light curves in the same transit events.
Two light curves from Gillon et al. (2012) are for 
the same transit event, i.e. epoch 38, 
and another two from Ricci et al. (2015) are for epoch 1469.
 
A Gaussian prior centered on the values 
of quadratic limb darkening coefficients
with certain $\sigma$ are set for TAP runs. 
The quadratic limb darkening coefficients and $\sigma$ 
for $I+z$ and Gunn-$r'$ filters 
are set as the values in Gillon et al. (2012),
and the one for white light curve follows the values used 
in Murgas et al. (2014).

For $i, I, J, R$, and $V$ filters, 
we linearly interpolate 
from Claret (2000, 2004) to the values effective temperature 
$T_{\rm eff}$ = 4400 K, log$g$ = 4.5 ${\rm cm/s^{2} }$, 
metallicity [Fe/H] = 0, and micro-turbulent velocity 
$V_{\rm t}$ = 0.5 ${\rm km/s}$ (Hellier et al. 2011).
In order to consider the possible small differences mentioned in 
Southworth (2008), a Gaussian prior centered 
on the theoretical values with $\sigma$ = 0.05 is set 
for our limb darkening coefficients $u_1$ and $u_2$ during 
TAP runs.
The details of parameter setting for TAP runs are listed in 
Table 3 and Table 4.

There are five chains in each of our TAP runs, and all of the chains 
are added together into the final results. The 15.9, 50.0 and 
84.1 percentile levels are recorded. The 50.0 percentile, i.e., median level, 
is used as the best value, and the other two percentile levels 
give the error bar.

The mid-transit time for the corresponding epoch
of each transit event is obtained.
In order to examine whether there is any outlier, these mid-transit times are
fitted by a linear function. It is found that the mid-transit time of 
epoch 1469 has the largest deviation and is more than 3$\sigma$ away 
from the linear function.  We thus remove two light curves of 
epoch 1469 from our data set and re-run TAP through the same procedure.   
We finally obtain the mid-transit time for the corresponding epoch
of each transit event, as those presented in Table 5. 
They will be used to establish a new ephemeris and study 
the transit timing variations in next section.
The results of inclination,
semi-major axis, and planet's radius are listed in Table 6.
These values are consistent with all those published in previous work.
For example, comparing with the results in Gillon et al.(2012) 
or Ricci et al (2015), our results 
are all extremely close to theirs, if error bars are considered. 
Our error bars are actually smaller than theirs. 
This shows that our analysis with more light curves gives 
stronger observational constraints. 

Moreover, the observational light curves and 
best fitting models of our own data are presented in Figure 1, 
where the points are observational data and solid curves are 
the best fitting models.
These eight light curves of our own work 
are available in a machine-readable form in 
the electronic version of Table 7.

\section{Transit Timing Variations}

\subsection{A New Ephemeris}

When all mid-transit times of 39 epochs in Table 5 
are considered, we obtain a new ephemeris 
by minimizing $\chi^2$ through 
fitting a linear function as
\beq
 T^C_{\rm m} (E) = T_0 + P E,
\eeq
where $T_0$ is a reference time, $E$ is an epoch 
(The transit observed in Hellier et al. 2011 is defined to be 
epoch $E=0$, and other transits' epochs are defined accordingly.), 
$P$ is the orbital period, and $T^C_{\rm m}(E)$ 
is the calculated mid-transit time at a given epoch $E$.

We find that
$T_{0} = 2455528.86860518\pm 0.00003632$ ($BJD_{TDB}$), 
$P = 0.81347392\pm 0.00000004$ (day).
The corresponding $\chi^2$ = 266.2076. 
Because the degree of freedom is 37, the reduced $\chi^2$, 
$\chi^2_{red}(37)$ = 7.1948.
Using this new ephemeris, 
the $O-C$ diagram is presented as the data points in Figure 2.
The large value of reduced $\chi^2$ of the linear fitting presented here 
indicates that a certain level of TTVs does exist.

\subsection{A Model of Orbital Decay}

Through the transit timing analysis, Blecic et al. (2014)
and Murgas et al. (2014) proposed a possible orbital decay for 
the planet WASP-43b. However, their transit data were up to about 
epoch 1000 only. It would be very interesting to see 
whether our newly observed data gives the transit timing 
with a trend of orbital decay.

Assume the orbital period is $P_q$, and the predicted 
mid-transit time at epoch $E$ is  $T_{S}(E)$.
For convenience, the mid-transit time of epoch 0, $T_S(0)$, is set to be zero, 
so the mid-transit time of epoch 1
is $T_S(1)=P_q$, and the elapsed time 
$\delta t_1= P_q$. If there is a small amount of 
period changing $\delta P$ from time $t= T_S(1)$ to 
$t=T_S(2)$, the elapsed time is $\delta t_2= P_q + \delta P$. 
Suppose there is a further period changing with $\delta P$
from time $t= T_S(2)$ to $t= T_S(3)$, 
so the elapsed time $\delta t_3= P_q + 2\delta P$.
Following this continuous period decay, we
have  $\delta t_i= P_q + (i-1)\delta P$,
where $i=1, 2, ...,(E-1), E$.
Summing up all the above $\delta t_i$, we obtain
$T_{S}(E) = E P_q + [E(E-1)/2]\delta P $.

Therefore, 
as in Blecic et al. (2014), a model of orbital decay can be obtained
by minimizing $\chi^2$ through 
fitting a function as\\
\begin{equation}
 T_{S}(E) = T_{q0} + P_q E + \delta P \frac{E(E-1)}{2}
\end{equation}
where $T_{q0}$ is a reference time, $E$ is an epoch, 
$P_{q}$ is the orbital period, $\delta P$ 
is the amount of period changing between each mid-transit time 
starting from $t=T_S(1)$. 

When only the data of earlier work 
with transits before epoch 1100, i.e. 
Gillon et al. (2012), Chen et al. (2014),  
Maciejewski et al. (2013), and Murgas et al. (2014), 
are considered, we have 
$T_{q0} = 2455528.86809115\pm 0.00006471$, 
$P_q = 0.81347925\pm 0.00000055$,
$\delta P = -1.03181346\times 10^{-8}\pm 0.10711789\times 10^{-8}$.
The corresponding $\chi^2$= 131.7672, and 
$\chi^2_{red}(24)= 5.4903$.
Using the above best-fitted parameters for 
$T_{S}(E)$  and the new ephemeris for $T^C_{\rm m}(E)$, 
the $ T_{S}(E)- T^C_{\rm m}(E)$ 
as a function of epoch $E$ is plotted as the dashed curve 
in the $O-C$ diagram, together with
data points as shown in Figure 2.
Both the units of $P_q$ and $\delta P$ are days,
and we obtain 
$dP/dt=\delta P/P_q$ = $-0.40027520\pm 0.04155436 $ sec/year.
We find that this result is consistent 
with the orbital decay rate stated in previous works. 

When all the data in Table 5 are considered, we obtain
$T_{q0} = 2455528.86851783\pm 0.00004318$, 
$P_q=0.81347448\pm 0.00000016$,
$\delta P = -7.45173434\times 10^{-10}\pm 1.98109164\times 10^{-10}$.
The corresponding $\chi^2_{red}(36)=7.0057$.
The larger value of reduced $\chi^2$ is due to the 
larger number of data points in this case.
Using the above best-fitted parameters for 
$T_{S}(E)$  and the new ephemeris for $T^C_{\rm m}(E)$, 
the $ T_{S}(E)- T^C_{\rm m}(E)$ 
as a function of epoch $E$ is plotted as the solid curve 
in the $O-C$ diagram, together with
data points as shown in Figure 2.
Comparing the solid curve with the dashed curve in Figure 2, 
it is obvious that the 
data points around epoch 1500 and epoch 1900 do not follow the 
dashed curve. That is, the newly obtained transits do not
follow the predicted transit timings in previous models. 

On the other hand, for the solid curve in Figure 2, the overall 
orbital decay rate is 
$dP/dt=\delta P/P_q$ = $-0.02890795\pm 0.00772547$ sec/year, which is 
one order smaller than the values in previous work.
Therefore, with our newly observed transits, 
we obtain a very different orbital decay rate. 
These results indicate that,  
if there is any orbital decay, the decay rate shall be much smaller 
than those values proposed in previous works.
This slower orbital decay rate leads to a new estimate of 
the stellar tidal dissipation factor $Q_{\ast}$.
Following the equation in Blecic et al. (2014),
we obtain a value of $Q_{\ast}$ about the order of $10^5$,
which is within the range of normally assumed theoretical value
from $10^5$ to $10^{10}$.   

\subsection{The Frequency Analysis}

In order to search for possible periodicities of transit timing 
variations from the timing residuals, 
Lomb-Scargle normalized periodogram (Press \& Rybicki 1989) 
is used. Figure 3 shows the resulting spectral 
power as a function of frequencies. 
The false-alarm probability of the largest power 
of frequencies is 0.20, which is very far from the usual 
thresholds 0.05 or 0.01 for a confirmed frequency. 
Therefore, our results show that there is no evidence 
for periodic TTVs. 
 
\section{Concluding Remarks}

Employing telescopes at two observatories, 
we monitor the transits of exoplanet WASP-43b
and obtain eight new transit light curves.
Together with the light curves from published papers,
they are all further analyzed through the same procedure.
The transit timings are obtained, 
and a new ephemeris is established.  
The newly determined inclination $i=82.149^{+0.084}_{-0.086}$,
semi-major axis $a/R_{\ast}=4.837^{+0.021}_{-0.022}$, 
and planet's radius $R_{\rm p}/R_{\ast}=0.15929^{+0.00045}_{-0.00045}$ 
are all consistent with previous work.

Our results reconfirm that a certain level of 
TTVs does exist, which is the same as what was claimed 
in Blecic et al.(2014) and Murgas et al.(2014) previously.
However, the results here show that the transit timings of new data 
do not follow the fast trend of the orbital decay suggested
in Blecic et al. (2014) and Murgas et al. (2014).
Our results lead to an orbital decay rate 
$dP/dt$ = $ -0.02890795\pm 0.00772547$ sec/year, which is
one order smaller than the previous values.
This slower rate corresponds to a larger
stellar tidal dissipation factor $Q_{\ast}$ 
in the range of normally assumed theoretical value.

On the other hand, 
the false-alarm probabilities in the frequency analysis indicate
that these TTVs are unlikely to be periodic.  
The TTVs we present here could be signals of a slow orbital decay.
 
We conclude that, in order to further investigate and understand 
this interesting system, both realistic theoretical modeling
and much more high-precision observations 
are desired in the future.

\section*{Acknowledgment}
We thank the anonymous referee for good suggestions 
which greatly improved this paper. We also thank the 
helpful communications with 
Gillon, M., Gazak, J. Z., Maciejewski, G., and Ngeow, C.-C..
This work is supported in part
by the Ministry of Science and Technology, Taiwan, 
under MOST 103-2112-M-007-020-MY3 and 
       NSC 100-2112-M-007-003-MY3.

\clearpage
\vskip 0.1truein
\center{Table 1: The log of observations of this work}
\begin{table}[ht]
\begin{center}
\begin{tabular}{ccccccc}
\hline
Run & UT Date  & Instrument & Filter & Interval (JD-2450000) &
Exposure & No. of
Images \\
\hline   
 1   & 2012 Mar 24 &   AZT-11   &   R    &  6011.212 - 6011.302  &    30
& 251 \\
 2   & 2014 Mar 12 &    P60     &   R    &  6728.698 - 6728.789  &    10
& 219 \\
 3   & 2014 Mar 16 &    P60     &   R    &  6732.767 - 6732.856  &    10
& 213 \\
 4   & 2014 Apr 07 &    P60     &   R    &  6754.731 - 6754.820  &    12
& 205 \\
 5   & 2014 Dec 24 &    P60     &   R    &  7015.864 - 7015.940  &    12
& 150 \\
 6   & 2015 Jan 06 &    P60     &   R    &  7028.864 - 7028.955  &    12
& 194 \\
 7   & 2015 Jan 15 &    P60     &   R    &  7037.815 - 7037.905  &    12
& 190 \\
 8   & 2015 Jan 19 &    P60     &   R    &  7041.882 - 7041.979  &    12
& 194 \\
\hline
\end{tabular}
\caption[Table 1]{The log of observations of this work.
For each run, the UT date,
instrument, filter, observational interval (JD-2450000),
exposure time (second), and the number of images
are listed.}
\end{center}
\end{table}

\clearpage
\vskip 0.1truein
\center{Table 2: The numbers of stars}
\begin{table}[ht]
\begin{center}
\begin{tabular}{ccccc}
\hline
 Run & No. of Brighter Stars & No. of Candidates & No. of Comparisons & oot rms \\
\hline
 1 &  5                &  3                &  1                 & 0.0045 \\
 2 &  18               &  3                &  3                 & 0.0028 \\
 3 &  12               &  4                &  2                 & 0.0041 \\
 4 &  24               &  5                &  2                 & 0.0020 \\
 5 &  5                &  2                &  2                 & 0.0029 \\
 6 &  11               &  2                &  2                 & 0.0035 \\
 7 &  16               &  3                &  3                 & 0.0042 \\
 8 &  13               & 10                &  6                 & 0.0027 \\
\hline
\end{tabular}
\caption[Table 2]{The number of stars in the process for choosing comparison
stars. The out-of-transit
root-mean-square of light curves are also listed.}
\end{center}
\end{table}

\clearpage
\vskip 0.1truein
\center{Table 3: The parameter setting}
\begin{table}[ht]
\begin{center}
\begin{tabular}{lll}
\hline
 Parameter & Initial Value         & During MCMC Chains \\
\hline
$P$(day) & 0.81347753        &  fixed \\
$i$(deg) & 82.33             & free, linked among all \\
 $a$/$R_{\ast}$         & 4.918  & free, linked among all  \\
 $R_{\rm p}$/$R_{\ast}$ & 0.15945 & free, linked among all \\
 $T_{\rm m }$         & set-by-eye & free, only linked if same transit events\\
 $u_1$             & Claret (2000,2004)    & a Gaussian prior, not linked \\
 $u_2$             & Claret (2000,2004)    & a Gaussian prior, not linked \\
 $e$                    & 0.0                   & fixed \\
 ${\omega}$             & 0.0                   & fixed \\

\hline
\end{tabular}
\caption[Table 3]{
The parameter setting. The initial values of $P, i, a/R_{\ast}, 
R_{\rm p}/R_{\ast}$  are adopted from Table 5 of Gillon et al.(2012).}
\end{center}
\end{table}

\clearpage
\vskip 0.1truein
\center{Table 4: The quadratic limb darkening coefficients}
\begin{table}[ht]
\begin{center}
\begin{tabular}{lcc}
\hline
 filter         & $u_1$                 & $u_2$ \\
\hline
 $^{a}I+z$      & $0.440\pm 0.035$    & $0.180\pm 0.025$   \\
 $^a$Gunn-$r'$  & $0.625\pm 0.015$    & $0.115\pm 0.010$   \\
 $^{b}$white    & $0.394\pm 0.087$    & $0.289\pm 0.119$   \\
 $^{c}i$        & $0.4767\pm 0.05$    & $0.2067\pm 0.05$   \\
 $^{c}I$        & $0.4401\pm 0.05$    & $0.2200\pm 0.05$   \\
 $^{c}J$        & $0.2560\pm 0.05$    & $0.2959\pm 0.05$   \\
 $^{c}R$        & $0.6012\pm 0.05$    & $0.1492\pm 0.05$   \\
 $^{c}V$        & $0.7598\pm 0.05$    & $0.0427\pm 0.05$   \\
 $^{d}$clear    & $0.6805\pm 0.05$    & $0.0960\pm 0.05$   \\
\hline
\end{tabular}
\caption[Table 4]{The quadratic limb darkening coefficients.\\
$^a$set as the values in Gillon et al. (2012)\\
$^b$set as the values in Murgas et al. (2014)\\
$^c$calculated for $T_{\rm eff} = 4400$ K, 
log$g$ = 4.5 ${\rm cm/s^{2} }$, [Fe/H] = 0, 
and $V_{\rm t}$ = 0.5 ${\rm km/s}$.\\
$^d$calculated as the average of those for $V$ and $R$ bands.
}
\end{center}
\end{table}

\clearpage
\vskip 0.1truein
\center{Table 5: The results of light-curve analysis for 
the mid-transit time}
\begin{table}[htb]\small
\begin{center}
\begin{tabular}{ccc}
\hline
Epoch  & Data Source &  $T_m$($BJD_{TDB}-2450000$) \\
\hline
11 & (a) & 5537.81659 $^{+0.00045}_{-0.00048}$ \\
22 & (a) & 5546.76493 $^{+0.00020}_{-0.00021}$ \\
27 & (a) & 5550.83218 $^{+0.00019}_{-0.00018}$ \\
38 & (a) & 5559.78048 $^{+0.00012}_{-0.00012}$ \\
43 & (a) & 5563.84773 $^{+0.00021}_{-0.00020}$ \\
49 & (a) & 5568.72833 $^{+0.00012}_{-0.00012}$ \\
59 & (a) & 5576.86368 $^{+0.00015}_{-0.00015}$ \\
65 & (a) & 5581.74392 $^{+0.00011}_{-0.00011}$ \\
70 & (a) & 5585.81297 $^{+0.00029}_{-0.00029}$ \\
76 & (a) & 5590.69256 $^{+0.00019}_{-0.00018}$ \\
87 & (a) & 5599.64047 $^{+0.00024}_{-0.00024}$ \\
97 & (a) & 5607.77505 $^{+0.00012}_{-0.00012}$ \\
124 &(a) & 5629.73981 $^{+0.00011}_{-0.00010}$ \\
140 &(a) & 5642.75474 $^{+0.00013}_{-0.00013}$ \\
141 &(a) & 5643.56875 $^{+0.00023}_{-0.00022}$ \\
152 &(a) & 5652.51574 $^{+0.00038}_{-0.00038}$ \\
168 &(a) & 5665.53206 $^{+0.00026}_{-0.00026}$ \\
173 &(a) & 5669.59939 $^{+0.00017}_{-0.00018}$ \\
189 &(a) & 5682.61543 $^{+0.00018}_{-0.00018}$ \\
200 &(a) & 5691.56374 $^{+0.00013}_{-0.00013}$ \\
211 &(a) & 5700.51237 $^{+0.00018}_{-0.00018}$ \\
243 &(a) & 5726.54407 $^{+0.00018}_{-0.00018}$ \\
499 &(b) & 5934.79276 $^{+0.00019}_{-0.00019}$ \\
543 &(c) & 5970.58598 $^{+0.00025}_{-0.00027}$ \\
593 &(f) & 6011.25910 $^{+0.00056}_{-0.00054}$ \\
950 &(d) & 6301.66872 $^{+0.00006}_{-0.00005}$ \\
1032&(c) & 6368.37476 $^{+0.00060}_{-0.00068}$ \\
1442&(e) & 6701.89857 $^{+0.00016}_{-0.00017}$ \\
1475&(f) & 6728.74255 $^{+0.00023}_{-0.00024}$ \\
1480&(f) & 6732.80936 $^{+0.00049}_{-0.00047}$ \\
1485&(e) & 6736.87743 $^{+0.00046}_{-0.00048}$ \\
1486&(e) & 6737.69125 $^{+0.00022}_{-0.00022}$ \\
1496&(e) & 6745.82472 $^{+0.00034}_{-0.00035}$ \\
1507&(f) & 6754.77378 $^{+0.00016}_{-0.00016}$ \\
1550&(e) & 6789.75311 $^{+0.00033}_{-0.00033}$ \\
1828&(f) & 7015.89837 $^{+0.00025}_{-0.00024}$ \\
1844&(f) & 7028.91466 $^{+0.00024}_{-0.00023}$ \\
1855&(f) & 7037.86178 $^{+0.00029}_{-0.00028}$ \\
1860&(f) & 7041.92985 $^{+0.00020}_{-0.00020}$ \\
\end{tabular}
\caption[Table 5]{The results of light-curve analysis for 
the mid-transit time $T_m$.\\
The epoch is the number of transit calculated from the first transit 
presented in Hellier et al. (2011). Data sources: (a) Gillon et al.(2012),
(b) Chen et al.(2014), (c) Maciejewski et al.(2013), (d) Murgas et al.(2014),   (e) Ricci et al.(2015), and (f) this work.
}
\end{center}
\end{table}
 
\clearpage
\vskip 0.1truein
\center{Table 6: The results of light-curve analysis for 
the inclination, semi-major axis, and planet's radius}
\begin{table}[htb]
\begin{center}
\begin{tabular}{cc}
\hline
Parameter              &  Value  \\
\hline
$i$                    & 82.149 $^{+0.084}_{-0.086}$ \\
$a$/$R_{\ast}$         & 4.837 $^{+0.021}_{-0.022}$ \\
$R_{\rm p}$/$R_{\ast}$ & 0.15929 $^{+0.00045}_{-0.00045}$ \\
\hline
\end{tabular}
\caption[Table 6]{The results of light-curve analysis for the
inclination $i$, semi-major axis $a$/$R_{\ast}$, 
and planet's radius $R_{\rm p}$/$R_{\ast}$.
}
\end{center}
\end{table}

\clearpage
\vskip 0.1truein
\center{Table 7: The photometric light-curve data of this work}
\begin{table}[h]
\begin{center}
\begin{tabular}{cccc}
\hline
Run & Epoch & TDB-based BJD & Relative Flux \\
\hline
1 &   593   &  2456011.21824144    &  0.999658 \\
  &         &  2456011.21859823    &  0.997367 \\
  &         &  2456011.21895503    &  0.997222 \\
\hline
2 & 1475  &  2456728.70459992  &  1.005345 \\
  &       &  2456728.70501139  &  1.002147 \\
  &       &  2456728.70542342  &  1.001576 \\
\hline
3 & 1480  & 2456732.77312609   &  0.999979 \\
  &       & 2456732.77353873   &  1.004087 \\
  &       & 2456732.77395244   &  1.002193 \\
\hline
\end{tabular}
\caption[Table 7]{The photometric light-curve data of this work.
This table is available in its entirety in the on-line journal.
A portion is shown here for guidance.}
\end{center}
\end{table}

\clearpage
\begin{figure}
\includegraphics[angle=0,scale=.70]{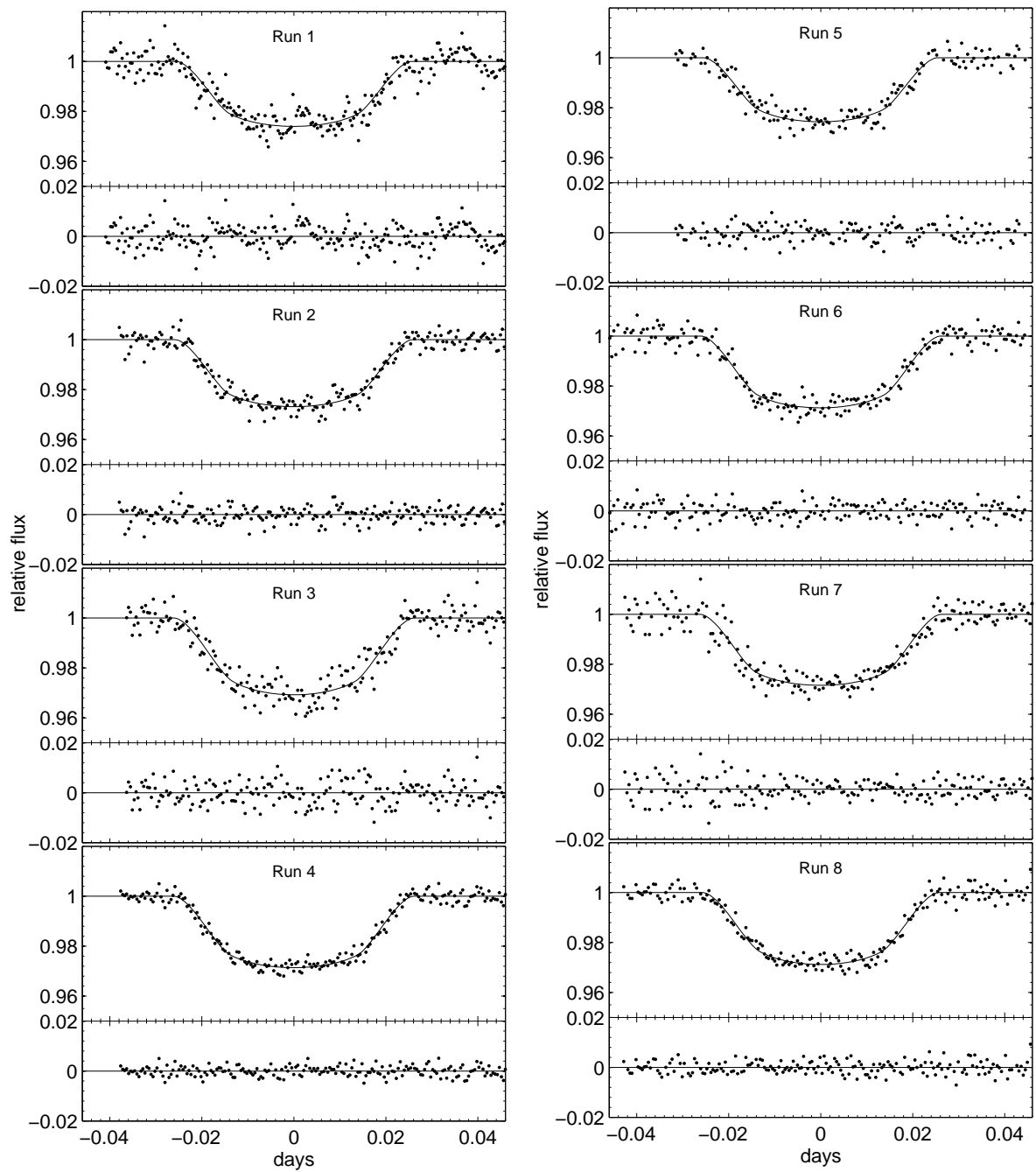}
\caption{The normalized relative flux as a function of
the time (the offset from mid-transit time and in TDB-based BJD)
of eight transit light curves of this work: points are the data and
curves are models. The corresponding residuals are shown at the
bottom of light curves. 
}
 \label {fig1}
 \end{figure}

\clearpage
\begin{figure}
\includegraphics[angle=0,scale=0.8]{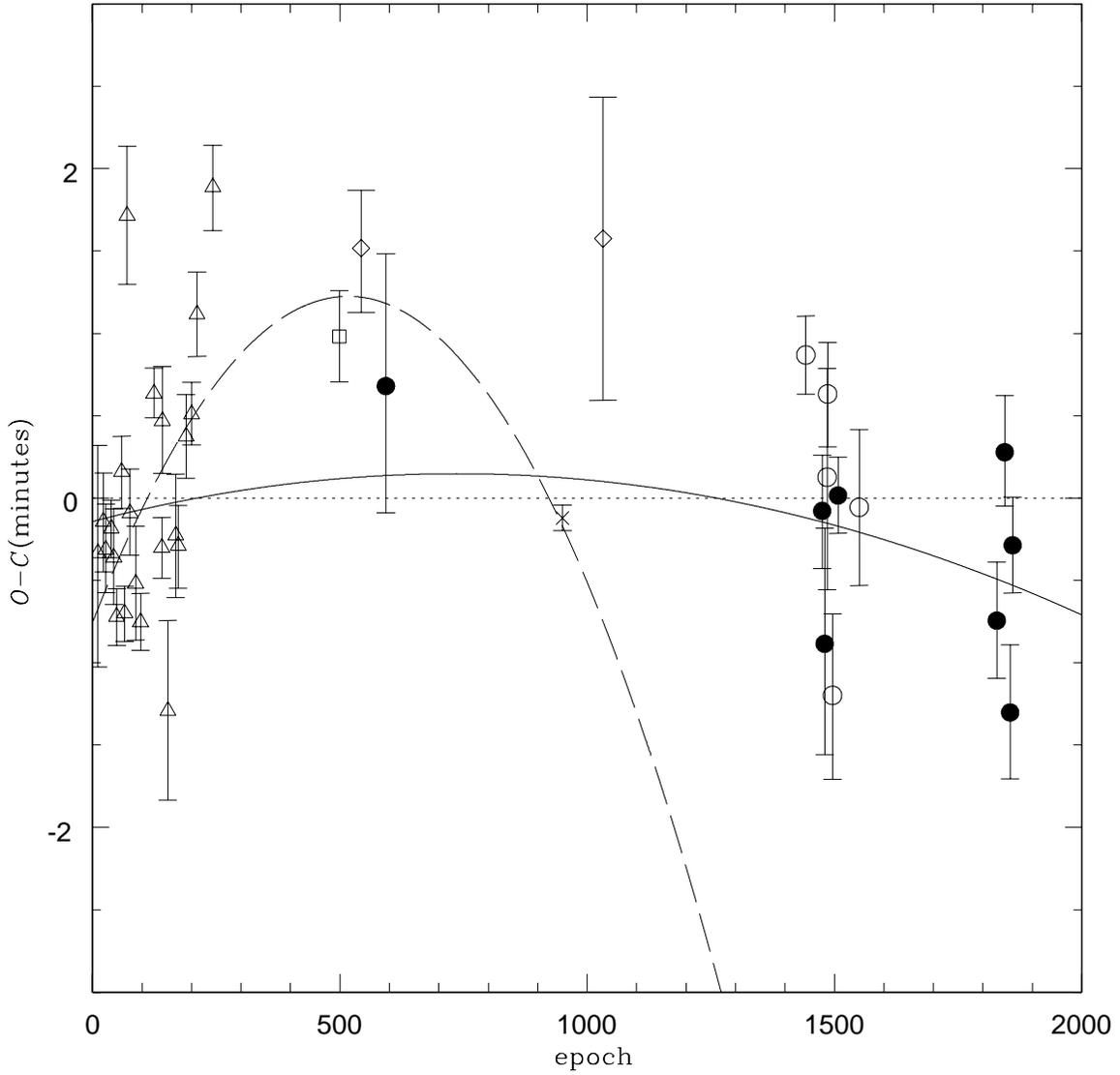}
\caption{The $O-C$ diagram. 
The filled circles are for our work. The triangles are 
for the data from Gillon et al. (2012), the square is for
the data from Chen et al. (2014), 
the diamonds are for the data from Maciejewski et al. (2013),
the cross is for the 
data from Murgas et al. (2014), and the open circles are 
for the data from Ricci et al. (2015). 
The dashed curve is the model determined by 
fitting with those data before epoch 1100 only. 
The solid curve is the model determined by 
fitting with all data.
}
 \label {fig2}
 \end{figure}

\clearpage
\begin{figure}
\includegraphics[angle=0,scale=.80]{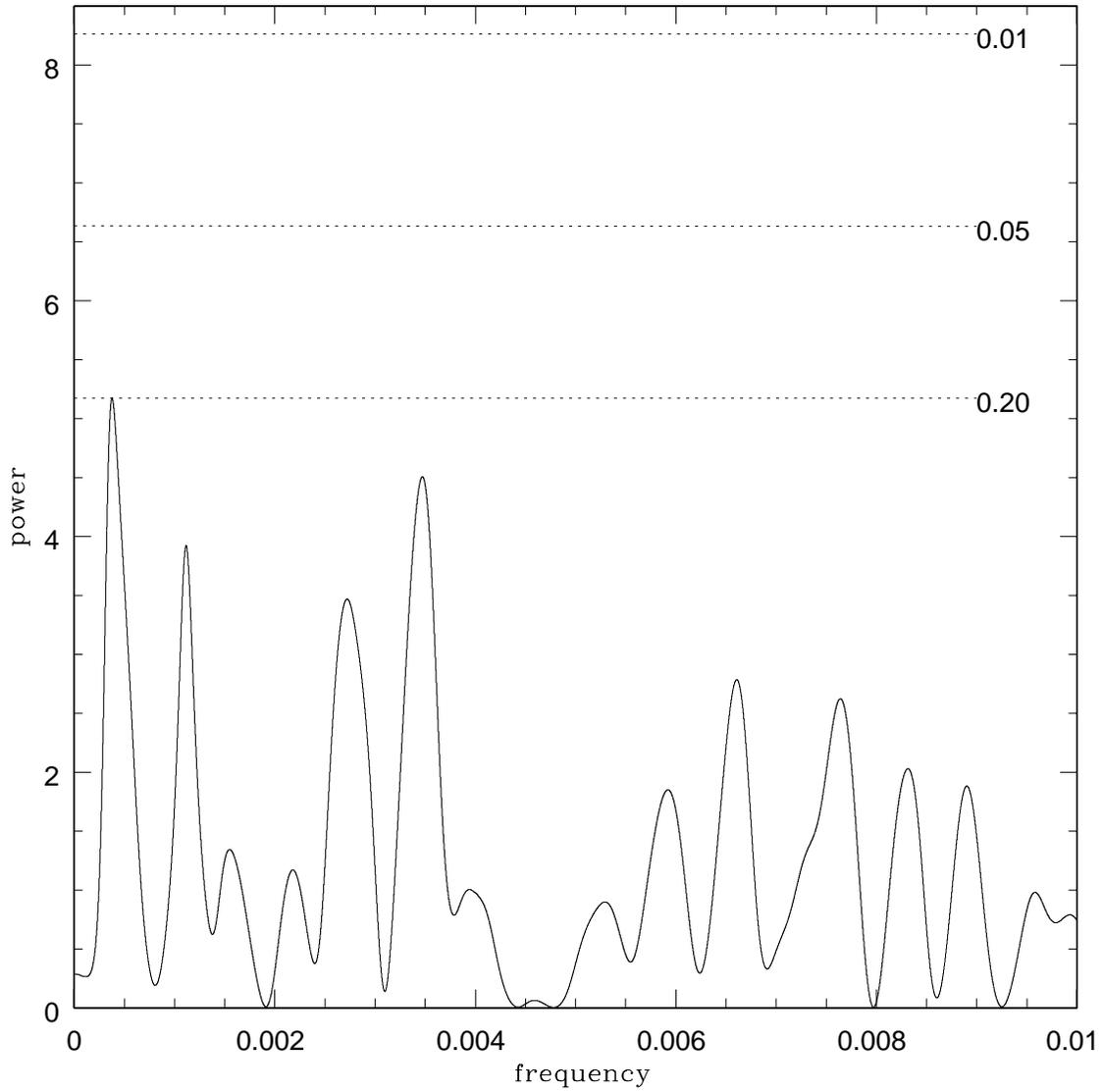}
\caption{The spectral power as a function of frequencies for the data points 
shown in Figure 2. 
 The false-alarm probability of the largest power 
of frequencies is 0.20 and shown 
as the bottom dotted line. The middle dotted line shows 0.05,
and the top dotted line shows 0.01 false-alarm probability.
}
\label {fig3}
\end{figure}

\end{document}